\documentclass[11pt,twocolumn]{article}
\usepackage{pslatex, amssymb, amsmath}
\usepackage{graphicx}
\usepackage{epstopdf}
\usepackage{algorithmic,algorithm}
\usepackage{subcaption}
\usepackage[latin9]{inputenc}
\usepackage{array}
\usepackage{float}
\usepackage{multirow}
\def\abstract{
\typeout{Abstract}
 {\bf Abstract} 
}

\begin{document}
\title{DEVELOPMENT OF C-MEANS CLUSTERING BASED ADAPTIVE FUZZY CONTROLLER FOR A FLAPPING WING MICRO AIR VEHICLE}
\date{}
\author{Md Meftahul Ferdaus, Sreenatha G. Anavatti, and Matthew A. Garratt \\ School of Engineering and Information Technolog, \\University of New South Wales at the Australian Defence Force Academy,\\ Canberra, ACT 2612, Australia
        \and Mahardhika Pratama\\ School of Computer Science and Engineering,\\ Nanyang Technological University,\\Singapore 639798, Singapore}

\maketitle
\begin{abstract}
Advanced and accurate modelling of a Flapping Wing Micro Air Vehicle (FW MAV) and its control is one of the recent research topics related to the field of autonomous Unmanned Aerial Vehicles (UAVs). In this work, a four wing Nature-inspired (NI) FW MAV is modeled and controlled inspiring by its advanced features like quick flight, vertical take-off and landing, hovering, and fast turn, and enhanced manoeuvrability when contrasted with comparable-sized fixed and rotary wing UAVs. The Fuzzy C-Means (FCM) clustering algorithm is utilized to demonstrate the NIFW MAV model, which has points of interest over first principle based modelling since it does not depend on the system dynamics, rather based on data and can incorporate various uncertainties like sensor error. The same clustering strategy is used to develop an adaptive fuzzy controller. The controller is then utilized to control the altitude of the NIFW MAV, that can adapt with environmental disturbances by tuning the antecedent and consequent parameters of the fuzzy system. 
\end{abstract}

\section{Introduction}
Nowadays, the application of autonomous systems in civil and military sector has increased significantly due to the advancement in control theory and power electronics devices. Among various autonomous systems, significant effort is invested in modelling and controlling Micro Air Vehicles (MAVs), and this is one of the latest research topics in the field of autonomous Unmanned Aerial Vehicles (UAVs). By definition MAVs usually have a maximum dimension of 150 mm, their size can be similar to a small bird, and they have a flight velocity of 10-20 $ms^{-1}$. Among these MAVs, very recently, nature-inspired flapping wing (NIFW) MAVs are becoming popular among researchers due to developments in micro-manufacturing techniques which have made easily realisable. NIFW MAVs are smaller in size and requires comparatively lower power than their fixed wings counterpart. The smaller size also provides them the capability to perform at lower Reynolds numbers which cannot be obtained from rotary wing UAVs. Furthermore, these BIFW MAVs are able to facilitate huge range of vital manoeuvres like vertical take-off and landing, gliding, roll banking, backward and sideways flying, which are not possible for similar sized fixed or rotary wing UAVs. Besides, BIFW MAVs have impressive potential in generating rapid acceleration during manoeuvres. The major benefits and feasibility of utilizing NIFW as MAV are described in \cite{ellington1999novel}. These huge benefits of NIFW MAVs over other fixed and rotary wing UAVs have made them worthy of investigation. 

The flight dynamics of NIFW MAVs, whether bird inspired or insect-inspired is more complex than their rotary or fixed-wing counterparts, since the flight solely depends on the beating motion of the flapping wings. Therefore, researchers have investigated the flight dynamics of various flapping wing creatures in the last two decades \cite{shyy2007aerodynamics,shyy2010recent,tennekes2009simple,willmott1997mechanics,sane2003aerodynamics,lin2017optimal,nogar2017control}. By analysing various features of nature-inspired flapping flight, the emphasis on developing NIFW MAVs is increasing in recent times \cite{zhang2017resonance,zhang2017review,oppenheimer2017wing}. 

Among different flying insects, dragonflies are one of the oldest with preferred mobility than most other insects as portrayed in \cite{couceiro2010modeling,sun2010coupled,okamoto1996aerodynamic,sudo1999study}. A dragonfly has four wings with the the ability of quick flight, hovering, and fast moves. In this manner, specialists are attempting to develop Dragonfly liked FW MAV (DLFW MAV) model and enhance their control precision. Linear and non-linear dynamics of a DL MAV is developed in \cite{jang2003longitudinal} from flight test information. Besides, a sliding mode control theory based adaptive controller was proposed in their work to stabilize the longitudinal dynamics. However, rather than flapping wing, their developed MAV model was a fixed wing one. A self-learning wing actuation system around a system of bearings for Dragonfly-inspired MAV (DI-MAV) was fabricated in \cite{kok2016design}. They have used three solenoids to generate three degrees of freedom motion from a wing, where the solenoids are controlled by motor driver. The drivers get sinusoidal signals from a workstation computer. A numerical model of a dragonfly robot is developed in \cite{couceiro2010modeling}. The model was tested with a ordinary PID and fractional order PID control algorithms in simulation. Their simulated model was able to mimic the kinematics and dynamics of a dragonfly. An insect-based FW MAV was fabricated in \cite{nguyen2014design}, which consisted of two fixed and two flapping wings. Their FW MAV can generate adequate vertical thrust to lift-off a weight of 14.76 grams at 10 Hz frequency. The maximum possible flapping frequency was 12.4 Hz, which produced an average vertical thrust to lift-off a weight of nearby 24 grams at an applied voltage of 3.7 V. Their FW-MAV exhibited a fruitful free flight with a decent control accuracy. A flight controller by utilizing  a Linear Quadratic Regulator (LQR) technique was developed in \cite{du2016application} for DLFW MAV. Their DLFW MAV model was linearised  to fit with the LQR flight controller. Besides, they have used an iterative learning based tuner to tune the input weighting matrix of LQR to deal with un-modelled parameters. To summarize, up to this point most of the strategies to model and control the FW MAV depend on first principle procedures, the exact numerical model is a need to manage their performance. But, the FW MAVs are profoundly nonlinear and over-actuated systems. They may contain different vulnerabilities. An exact numerical model of FW MAVs considering these features is challenging to achieve. A smart solution to these issues is the use of model free knowledge-based data-driven techniques.           

The data-driven modelling and control can play an important role in NIFW MAV system since they don't require any mathematical model. Some of the commonly used non-linear data-driven modelling and control techniques are: describing function method, block structured systems, fuzzy logic, neural networks, and Nonlinear Autoregressive Moving Average Model with Exogenous inputs (NARMAX methods). Among these techniques, fuzzy logic and neural network systems are promising since they demonstrate learning capability from a set of data and approximate reasoning trait of human beings which cope with the imprecision and uncertainty of the decision-making process. In recent times fuzzy logic and neural networks are employed to model and control various MAVs \cite{ferdaus2017McSIT2RFNN,Al-Mahasneh2017,ferdaus2017fuzzyclustering,ferdaus2017fuzzyclusteringFWMAV,ferdaus2017GENEFISFWMAV,ferdaus2017Gcontroller,almahasnehanavattigarratt2017}. 

A Spiking Neural Network (SNN) to control an FW MAV called RoboBee was proposed in \cite{clawson2016spiking}. In \cite{weng2013micro}, a Neural Immunology Network was proposed, which is inspired by the memory and immune system. Their controller can control the motion of FW MAVs by considering various system nonlinearities. Besides, their control method can deal effectively with external perturbations  and parameter variations since they do not depend on precise dynamics model. Laurent et al. A dynamic recurrent neural network was developed in \cite{laurentevolution}, which can produce oscillating behaviour. They had employed  ModNet encoding based evolutionary algorithm to evolve both the structure and weights of the neural network. Their neuro-controller can tune the dihedral, the twist and the sweep of the wings at every moment to balance the FW MAV body by generating sufficient lift and traction. A direct adaptive (DA) and hybrid adaptive fuzzy controller (HAFC) was developed in \cite{couceiro2012hybrid} to control dragonfly-like FW MAV model by simulation. Better trajectory tracking performance is observed from the HAFC than the DAFC. 

Due to the successful implementation and evaluation of various neuro and fuzzy technique in FW MAV, in our work a FCM clustering based identification technique is utilized and an adaptive controller for a NIFW MAV is developed. The advantage of the FCM technique is its adaptation capability with the sudden change in environments.

\section{Fuzzy Modelling and Adaptive Control of Flapping Wing Micro-Air Vehicle}
The FW MAV used in our work is a simulated nature-inspired insect robot with four wings. The development process of the NIFW MAV flight simulator is described in \cite{kok2015low}. From this flight simulator the data has been collected to develop the fuzzy based identification and adaptive controller. Due to the high cost and time consumption to develop and set-up experimental flight test, the utilization of such flight simulators are usual. In this flight simulator, the wing kinematics for a wing flapping in an inclined stroke plane are obtained from the derivation described in \cite{wang2004role}. The flapping angle ($\phi$) in the flapping profile of the FW MAV can be expressed as follows:
\begin{equation}\label{eq:1}
\phi(t)=\frac{\phi_{fa}}{2}\text{cos}(\pi ft)
\end{equation}
where $\phi_{fa}$ is the flapping amplitude in radian, $f$ is the flapping frequency in Hz, $t$ is the time is second. 
The angle of attack ($\alpha$) can be presented as:
\begin{equation}\label{eq:2}
\alpha=\alpha_{ma}-\alpha_0 \text{sin}(\omega dt+\psi)
\end{equation}
where $\alpha_{ma}$ is mean angle of attack in radian, $\alpha_0$ is amplitude of pitching oscillation in radian, $dt$ is time step in seconds, and $\psi$ is the phase difference between the pitching and plunging motion. All the four wings of the FW MAV follows the same flapping profile. 

\begin{figure}[!t]
	\centering
	\includegraphics[width=2.7in]{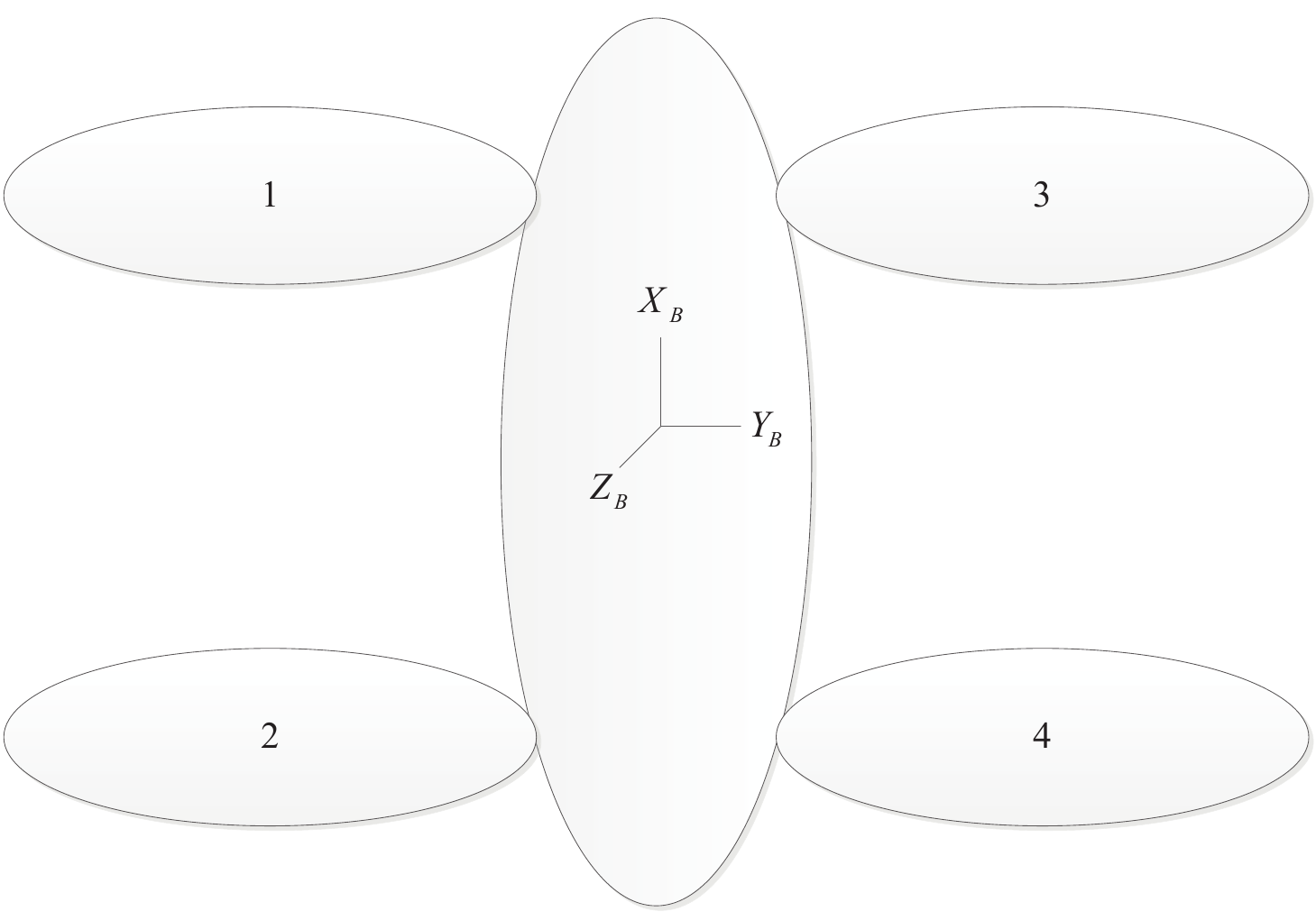}
	\caption{Body coordination of a NIFW MAV. Numbers indicate the actuator number}
	\label{fig:fwmav}
\end{figure}

In the simulator, each wing is controlled by an actuator. A symbolic diagram or body coordination of four wing NIFW MAV is exhibited in \figurename{~\ref{fig:fwmav}}.

Each actuator is controlled by eight (8) flapping parameters. A parametric analysis is accomplished to find the dominant flapping parameters. After a complete parametric analysis, it is observed that among the eight parameters the flapping amplitude is the dominant one to control the NIFW MAV. The effect of changing the flapping amplitude to some major manoeuvring of a NIFW MAV is summarized in \tablename{~\ref{tab:Effects of flapping amplitude}} .

\begin{table}[!t]
	\renewcommand{\arraystretch}{1.3}
	\caption{Effects of flapping amplitude in different manoeuvring of NIFW MAV}
	\label{tab:Effects of flapping amplitude}
	\centering
	\begin{tabular}{|c|p{2cm}|c|}
		\hline
		\bfseries Actuators & \bfseries  Flapping amplitude, $\phi_0$ (degree) & \bfseries Action \\
		\hline
		1, 2, 3, 4 & 90 & Take-off\\
		\hline
		1, 2 and 3, 4 & 90 and 60 & Roll-right\\
		\hline
		1, 2 and 3, 4 & 60 and 90 & Roll-left\\
		\hline
		1, 3 and 2, 4 & 90 and 60 & Pitch-up\\
		\hline
		1, 3 and 2, 4 & 60 and 90 & Pitch-down\\
		\hline
	\end{tabular}
\end{table} 

\subsection{Fuzzy Clustering Based Modelling of the NIFW MAV}
From the NIFW MAV flight simulator, the input-output data is collected to develop the data-driven model, where the four input datasets are the four flapping amplitudes applied to four actuators. The three-dimensional (3D) rotational velocities ($\omega_{bx}$, $\omega_{by}$, and $\omega_{bz}$) and translational velocities ($v_{bx}$,$v_{by}$, and $v_{bz}$) of the NIFW MAV body are six output datasets. From four inputs and six outputs, a Multi-Input Multi-output (MIMO) nonlinear NIFW MAV model is developed using fuzzy C-means (FCM) clustering technique. In FCM, a data sample may exist to more than one clusters with a degree of belongingness that varies between 0 to 1, where the integration of degrees of belongingness of a data sample to all groups is always one as expressed below:
\begin{equation}\label{eq:3}
\sum_{i=1}^{c}\mu_{ij}=1,~~\forall j=1,...,n
\end{equation}
However, the FCM still requires a cost function to be minimized during partitioning the data set. The cost function in FCM can be expressed as follows:
\begin{equation}\label{eq:4}
J(Y,y_1,...,y_c)=\sum_{i=1}^{c}J_i=\sum_{i=1}^{c}\sum_{j=1}^{n}\mu_{ij}^m d_{ij}^2
\end{equation}
where $\mu_{ij}$ ranges between 0 to 1; $y_c$ is the cluster center of the $c-th$ cluster; $d_{ij}=|||y_i-x_j|$ is the Euclidian distance between the $i-th$ cluster centre and $j-th$ data point; $m\in[1,\infty)$ is a weighting exponent.

The two conditions to reach the minimum for Equation \eqref{eq:4} are as follows:
\begin{equation}\label{eq:5}
y_i=\frac{\sum\limits_{j=1}^{n}\mu_{ij}^m x_j}{\sum\limits_{j=1}^{n}\mu_{ij}^m}
\end{equation}
\begin{equation}\label{eq:6}
\mu_{ij}=\frac{1}{\sum\limits_{k=1}^{c}\big(\frac{d_{ij}}{d_{kj}}\big)^{2/(m-1)}}
\end{equation}

The FCM algorithm repeatedly performs through Equation \eqref{eq:5} and \eqref{eq:6} until no more improvement is observed. The efficiency of Equation \eqref{eq:5} and \eqref{eq:6} and convergence of the FCM algorithm is proven in \cite{bezdek1980convergence}.  

In batch training operation, the FCM algorithm to determine the cluster centres $y_i$ and the membership matrix $Y$ is presented in Algorithm~\ref{alg:fcm_alg}below.
\begin{algorithm}
	\renewcommand\thealgorithm{}
	\caption{FCM Algorithm}\label{alg:fcm_alg}
	\begin{algorithmic}[1]
		\renewcommand{\algorithmicrequire}{\textbf{Input:}}
		\renewcommand{\algorithmicensure}{\textbf{Output:}}
		\REQUIRE Input/target pair
		\ENSURE  Identified output
		\\ \textit{Initialisation} :
		\STATE Initialize the membership matrix Y with random values ranging between 0 to 1 such a way to satisfy the constraints in Equation \eqref{eq:3}. 
		\\ \textit{LOOP Process}
		\FOR {$i = 1$ to $c$}
		\STATE Calculate the cluster centres $y_i$ using Equation \eqref{eq:4}
		\IF {(Cost function$>$ Threshold)}
		\STATE ADD new rule
		\ELSE
		\STATE Identified output=evalfis(input(i))
		\ENDIF
		\ENDFOR
		\RETURN Identified output=evalfis(input(i))
	\end{algorithmic}
	\addtocounter{algorithm}{-1}
\end{algorithm}

In FCM, the performance has dependency on the initial membership matrix values, which suggest running the algorithm for few times. 

\subsection{Development of Adaptive Fuzzy Controller}
\begin{figure*}[!t]
	\centering
	\includegraphics[width=5.7in]{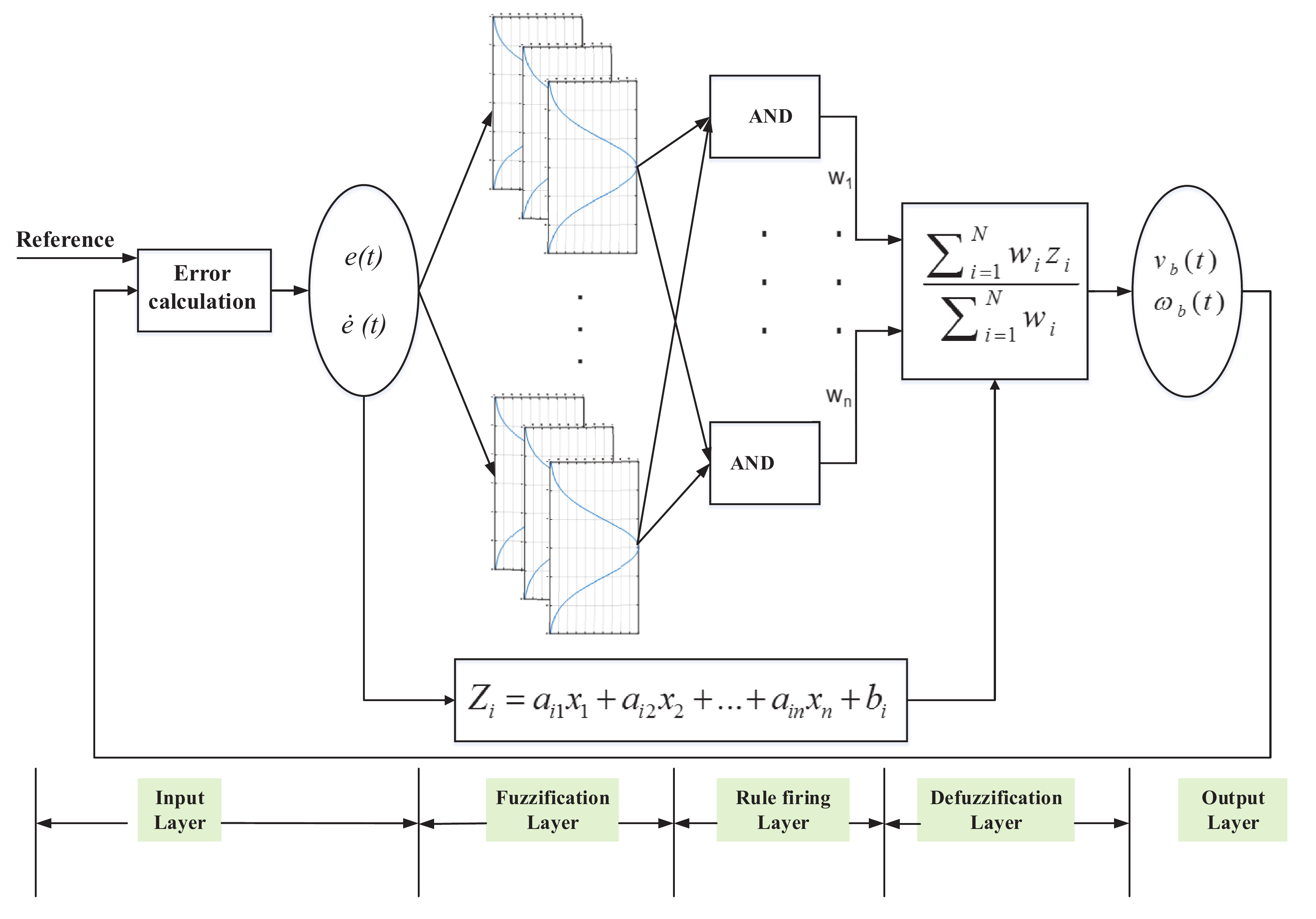}
	\caption{Closed-loop block diagram of FCM based fuzzy adaptive control system}
	\label{fig:controller}
\end{figure*}

The FCM clustering technique is utilized to develop an adaptive fuzzy controller. The closed-loop functional diagram of the fuzzy adaptive control system is exhibited in \figurename{~\ref{fig:controller}}. The difference between the Reference signal and identified model's output ($e(t)$) is one of the inputs to the controller, and the rate of change of that error ($de(t)/dt$) is another input to the controller, which is presented in the input layer of \figurename{~\ref{fig:controller}}. These crisp inputs are being fuzzified in the fuzzification layer, where Gaussian membership functions are utilized. To get the desired signal from the FW MAV model, the fuzzy controller alters the Gaussian membership functions width and centres by utilizing the FCM clustering technique, where the error signal $e(t)$ is utilized as a cost function for the controller. After this, the 'AND' operation i.e. the product of all membership functions are obtained. Finally, the output of the adaptive fuzzy controller is calculated in the output layer as follows:
\begin{equation}\label{eq:7}
y_{c_i}=\frac{\sum\limits_{i=1}^{N}w_{i}z_i}{\sum\limits_{i=1}^{N}w_{i}}
\end{equation}
The controller's output signal goes to the identified NIFW MAV model. Then the model's output is integrated to get the vertical altitude from velocity, and compared with the reference position. The controller tunes its parameter until the model output follows the reference signal, and consequently, the error signal ($e(t)$) becomes zero. In this FCM based adaptive fuzzy controller, five Gaussian membership functions are utilized.  

\section{Results and Discussion}                        
\begin{figure}[!t]
	\centering
	\includegraphics[width=3.5in]{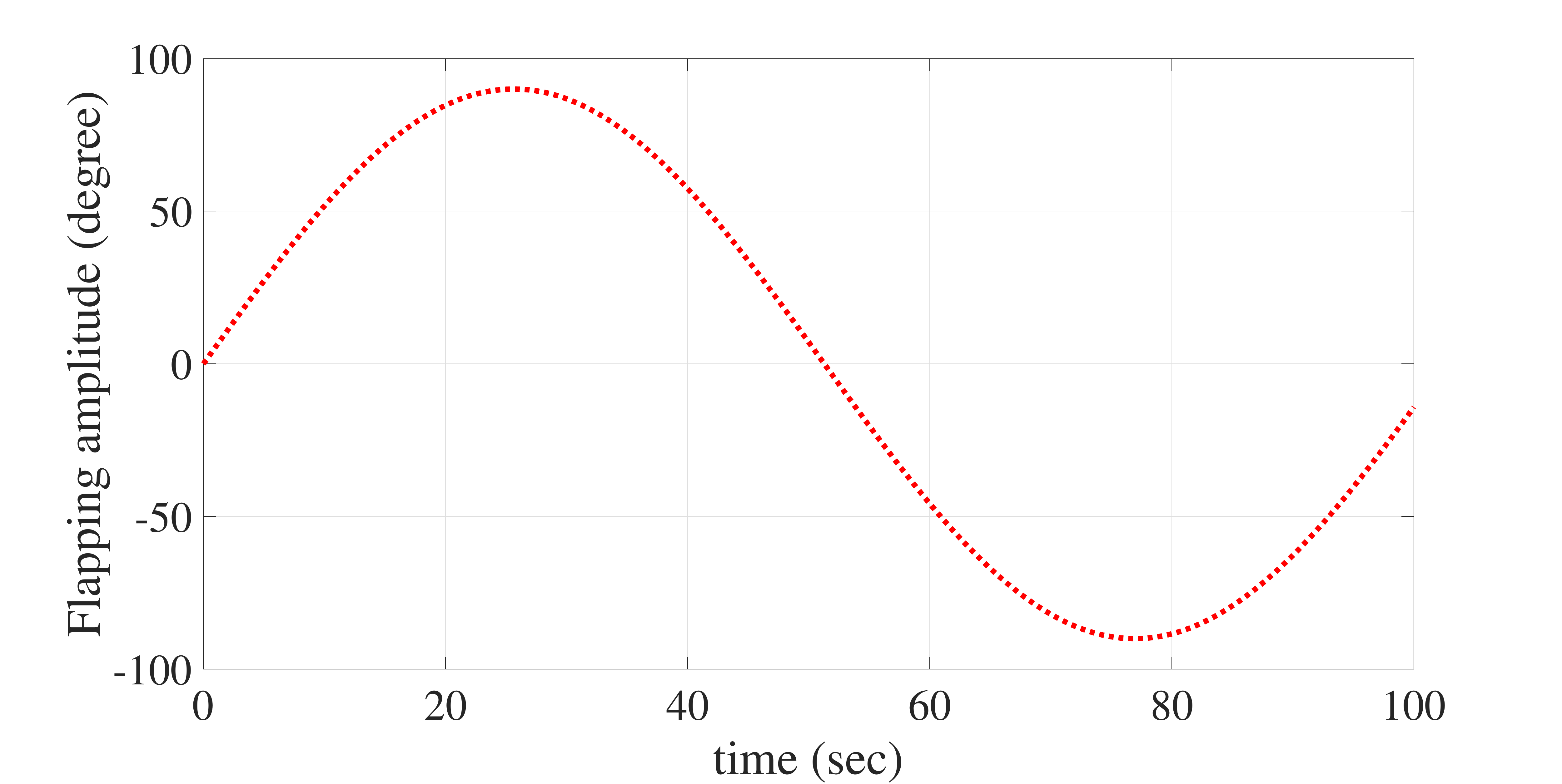}
	\caption{Input: Flapping amplitude}
	\label{fig:famp}
\end{figure}

\begin{figure}[!t]
	\begin{subfigure}{.5\textwidth}
		\centering
		\includegraphics[width=3.5in]{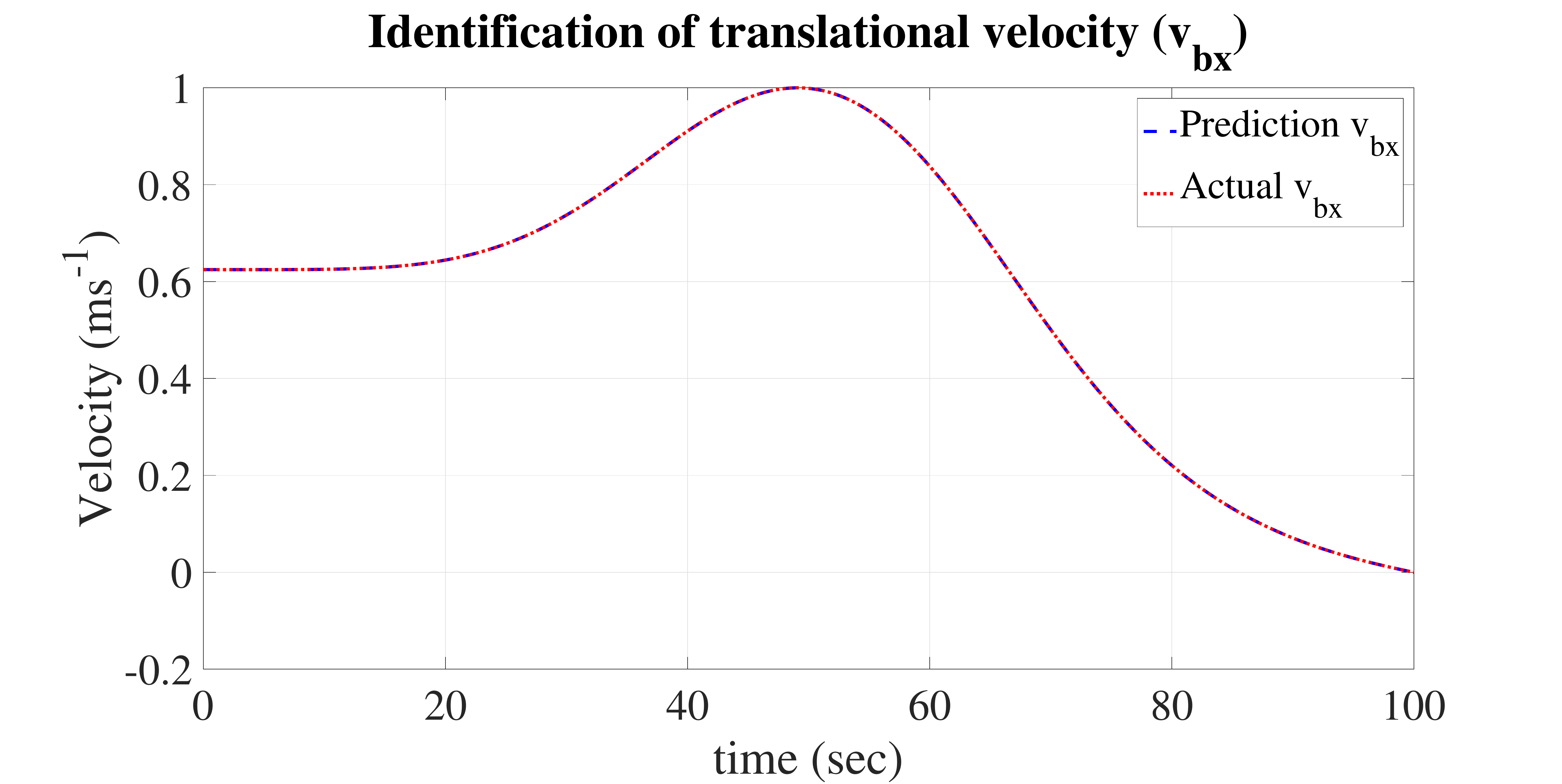}
		\caption{Translational velocity ($v_{bx}$)}
		\label{fig:sfig01}
	\end{subfigure}
	\begin{subfigure}{.5\textwidth}
		\centering
		\includegraphics[width=3.5in]{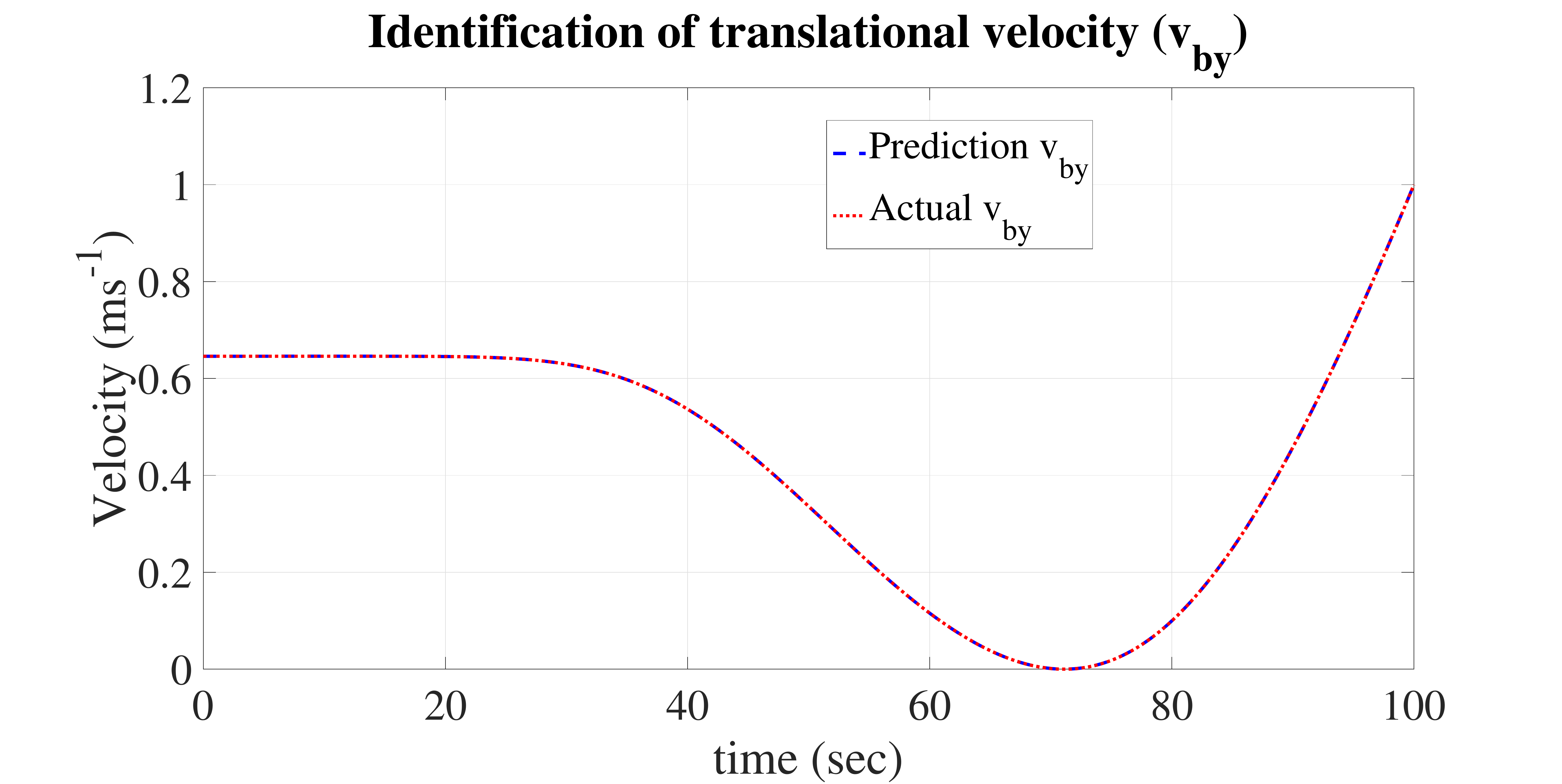}
		\caption{Translational velocity ($v_{by}$)}
		\label{fig:sfig02}
	\end{subfigure}\hspace{5mm}
	\begin{subfigure}{.5\textwidth}
		\centering
		\includegraphics[width=3.5in]{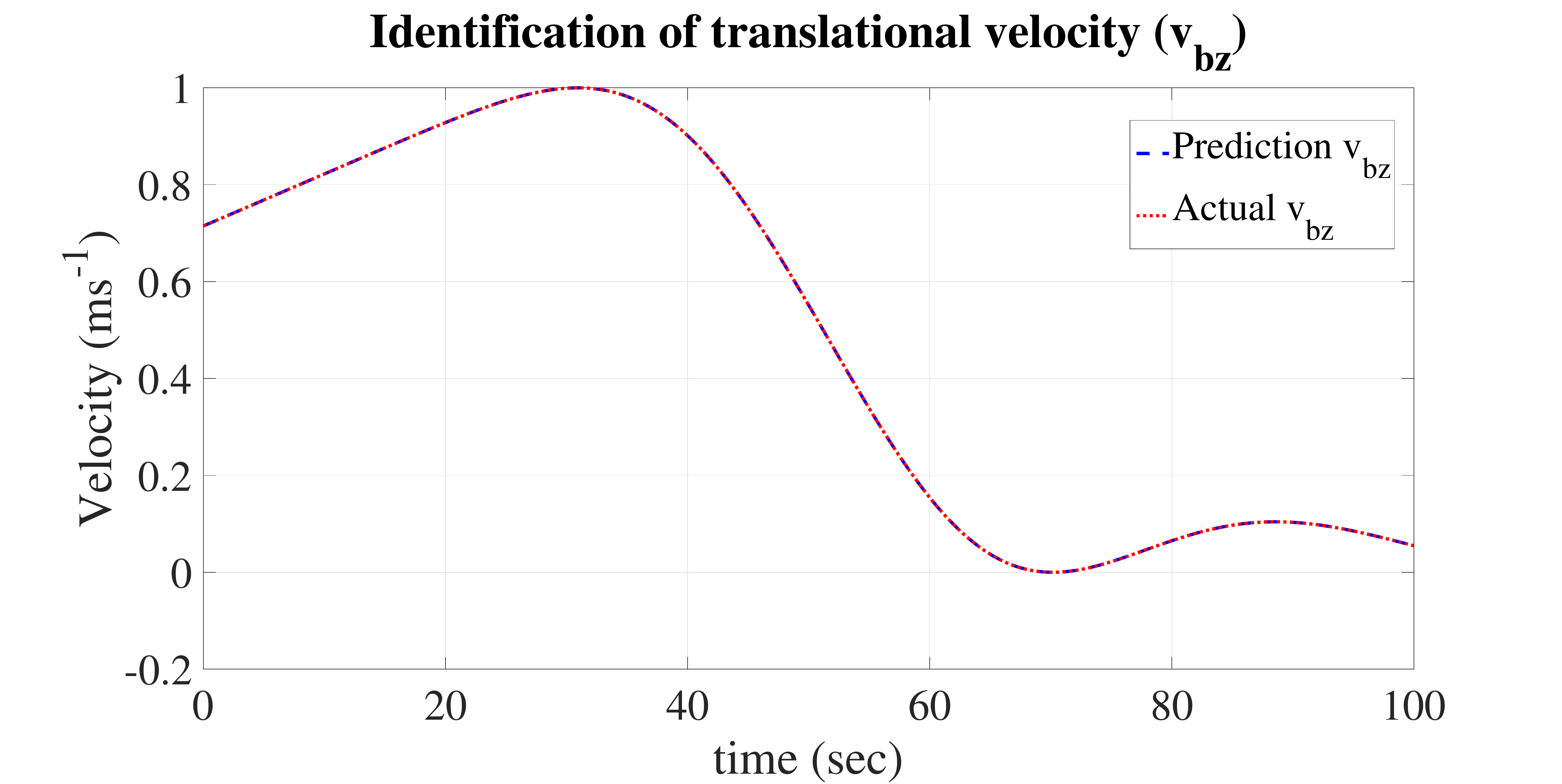}
		\caption{Translational velocity ($v_{bz}$)}
		\label{fig:sfig03}
	\end{subfigure}\hspace{5mm}
	\caption{Identification of FW MAVs translational velocity ($v_{bx}, v_{by}, v_{bz}$)}
	\label{fig:FWMAVident1}
\end{figure}

\begin{figure}[!t]
	\begin{subfigure}{.5\textwidth}
		\centering
		\includegraphics[width=3.5in]{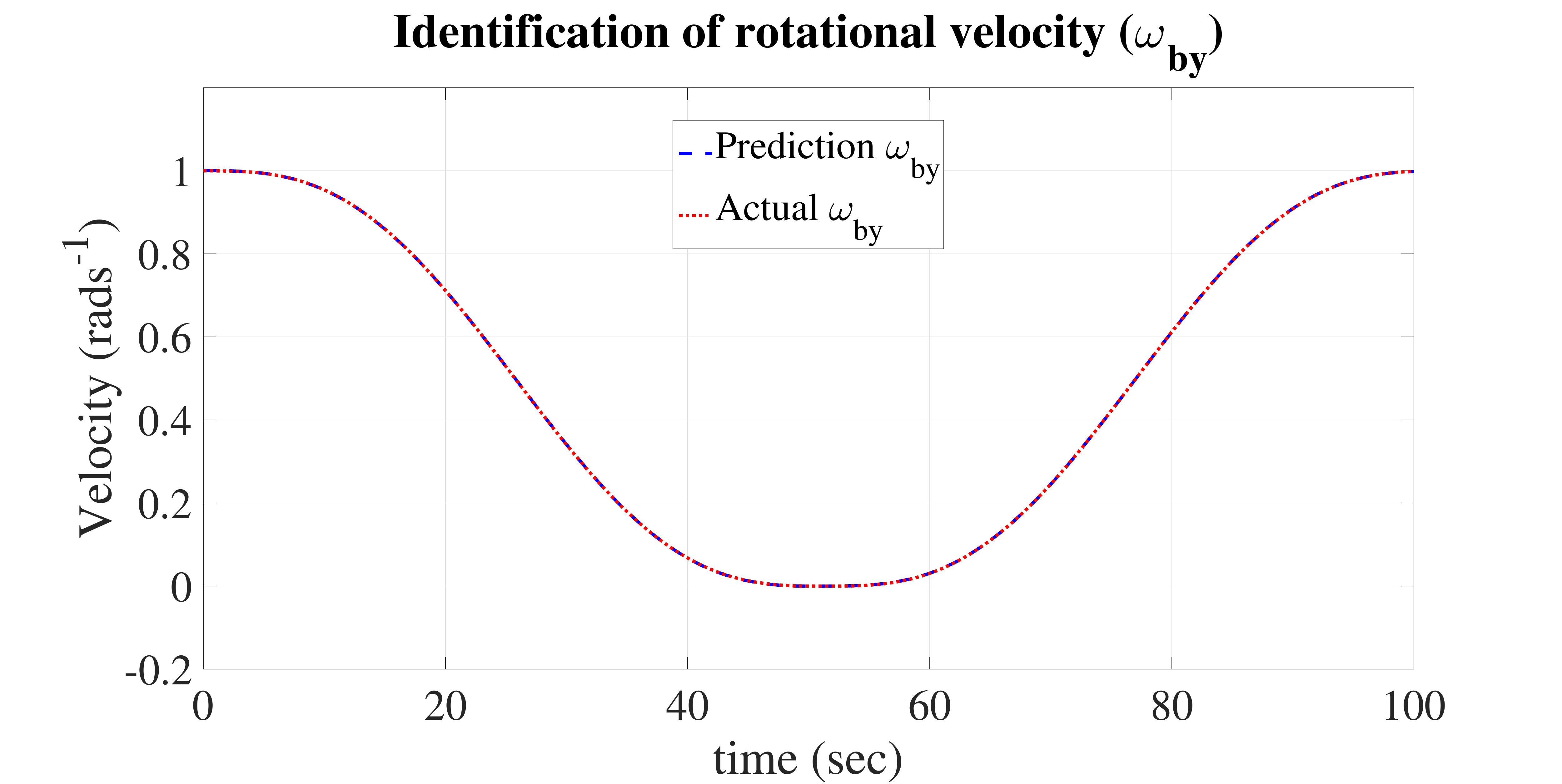}
		\caption{Rotational velocity ($\omega_{by}$)}
		\label{fig:sfig05}
	\end{subfigure}
	\begin{subfigure}{.5\textwidth}
		\centering
		\includegraphics[width=3.5in]{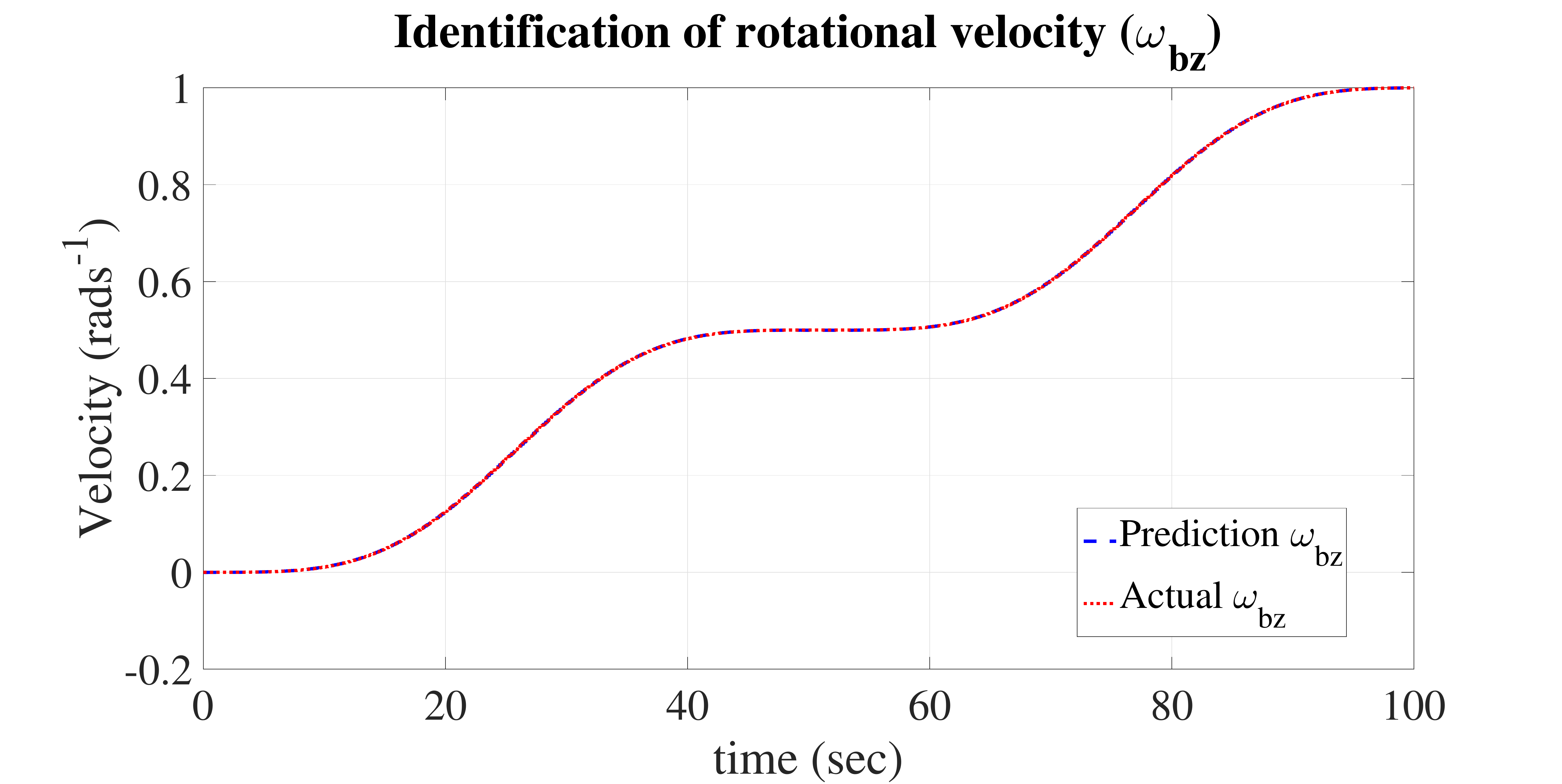}
		\caption{Rotational velocity ($\omega_{bz}$)}
		\label{fig:sfig06}
	\end{subfigure}
	\caption{Identification of FW MAVs translational velocity ($\omega_{bx}, \omega_{by}, \omega_{bz}$)}
	\label{fig:FWMAVident2}
\end{figure}

The data used for the MIMO nonlinear FW MAV system identification is based on a 100 seconds simulation in Simulink with a time step of 0.01 seconds. In a physical NIFW MAV model, they can change their flapping amplitude within a certain range, which is between $-90^{\circ}$ and $90^{\circ}$. Therefore, a sinusoidal flapping amplitude varying between $-90^{\circ}$ and $90^{\circ}$ is applied to all four actuators of the four wings of NIFW MAV as shown in \figurename{~\ref{fig:famp}}, which helps the FCM clustering based MAV model to get the input datasets within the maximum range. Besides, they are smooth in the way of changing, which can easily be recognized by the FCM clustering technique. In this technique, Takagi-Sugeno (TS) fuzzy model with three (3) Gaussian membership function is utilized. From \figurename{~\ref{fig:FWMAVident1}} and \figurename{~\ref{fig:FWMAVident2}} it is clearly observed that all the translation velocities ($v_{bx}$,$v_{by}$, and $v_{bz}$), and the rotational velocities ($\omega_{bx}$, $\omega_{by}$, and $\omega_{bz}$) are identified precisely. 

\begin{figure}[!t]
	\centering
	\includegraphics[width=3.5in]{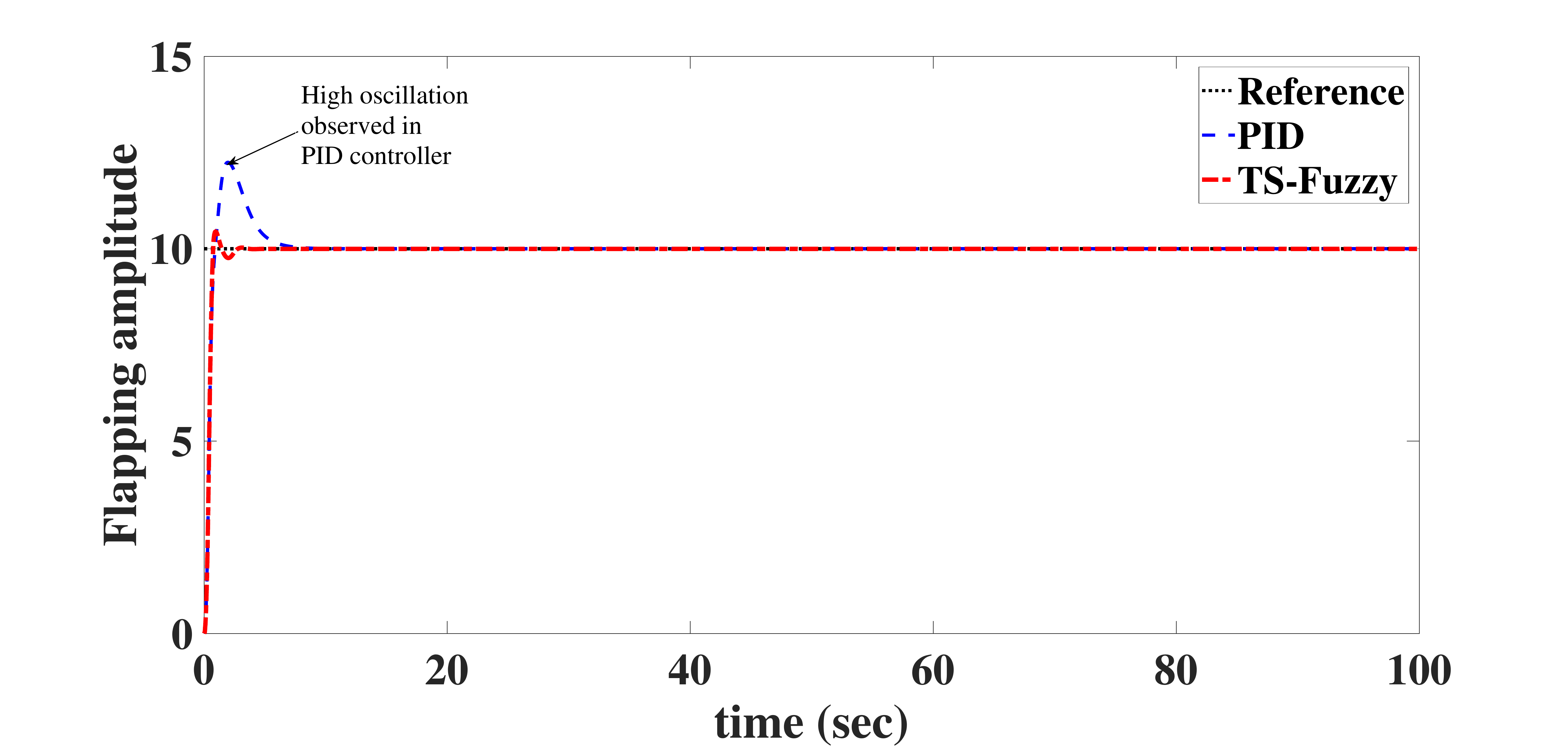}
	\caption{Altitude tracking performance of NIFW MAV controllers for a constant height trajectory}
	\label{fig:cons_height}
\end{figure}

\begin{figure}[!t]
	\centering
	\includegraphics[width=3.5in]{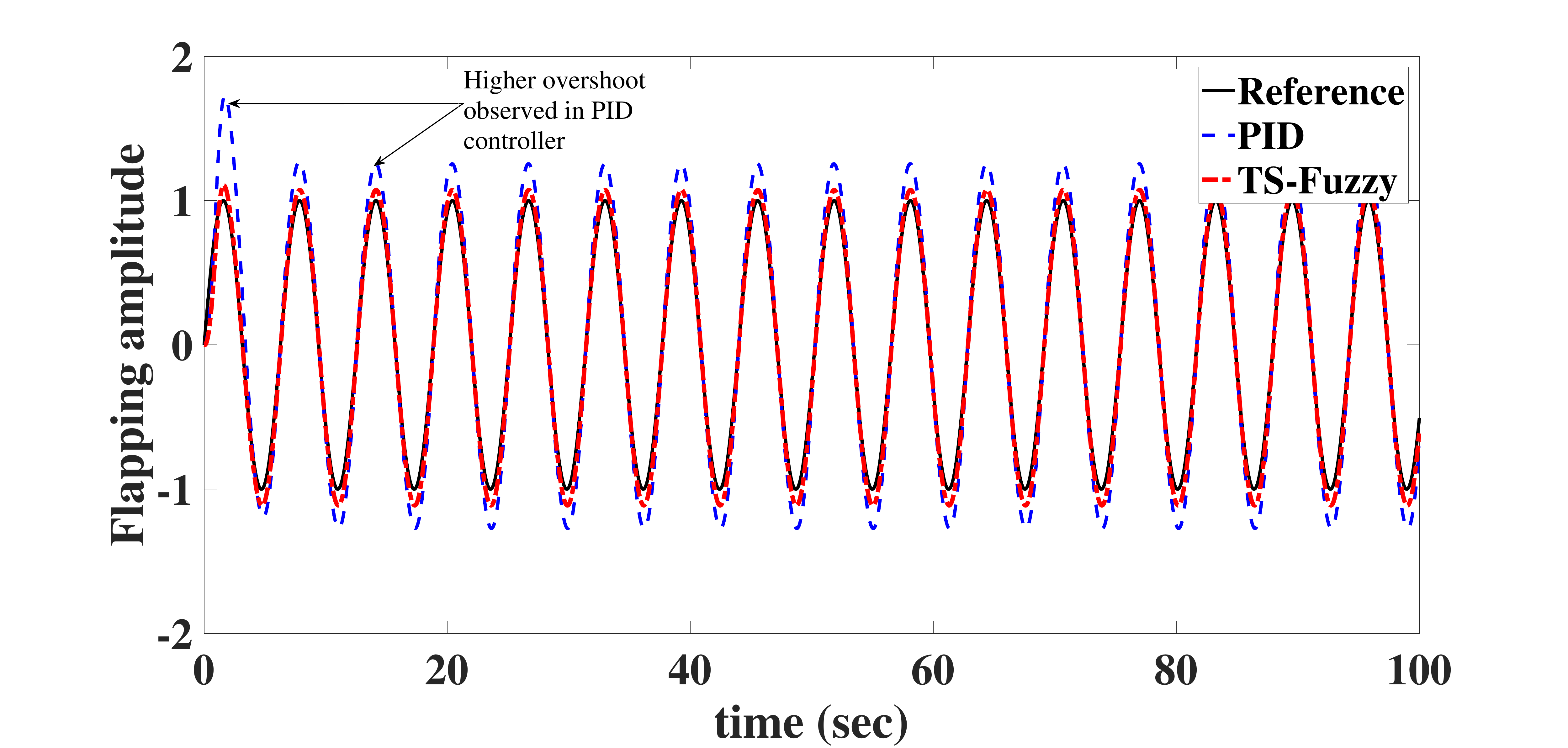}
	\caption{Altitude tracking performance of NIFW MAV controllers for a sinusoidal trajectory}
	\label{fig:sinusoidal}
\end{figure}

\begin{figure}[!t]
	\centering
	\includegraphics[width=3.5in]{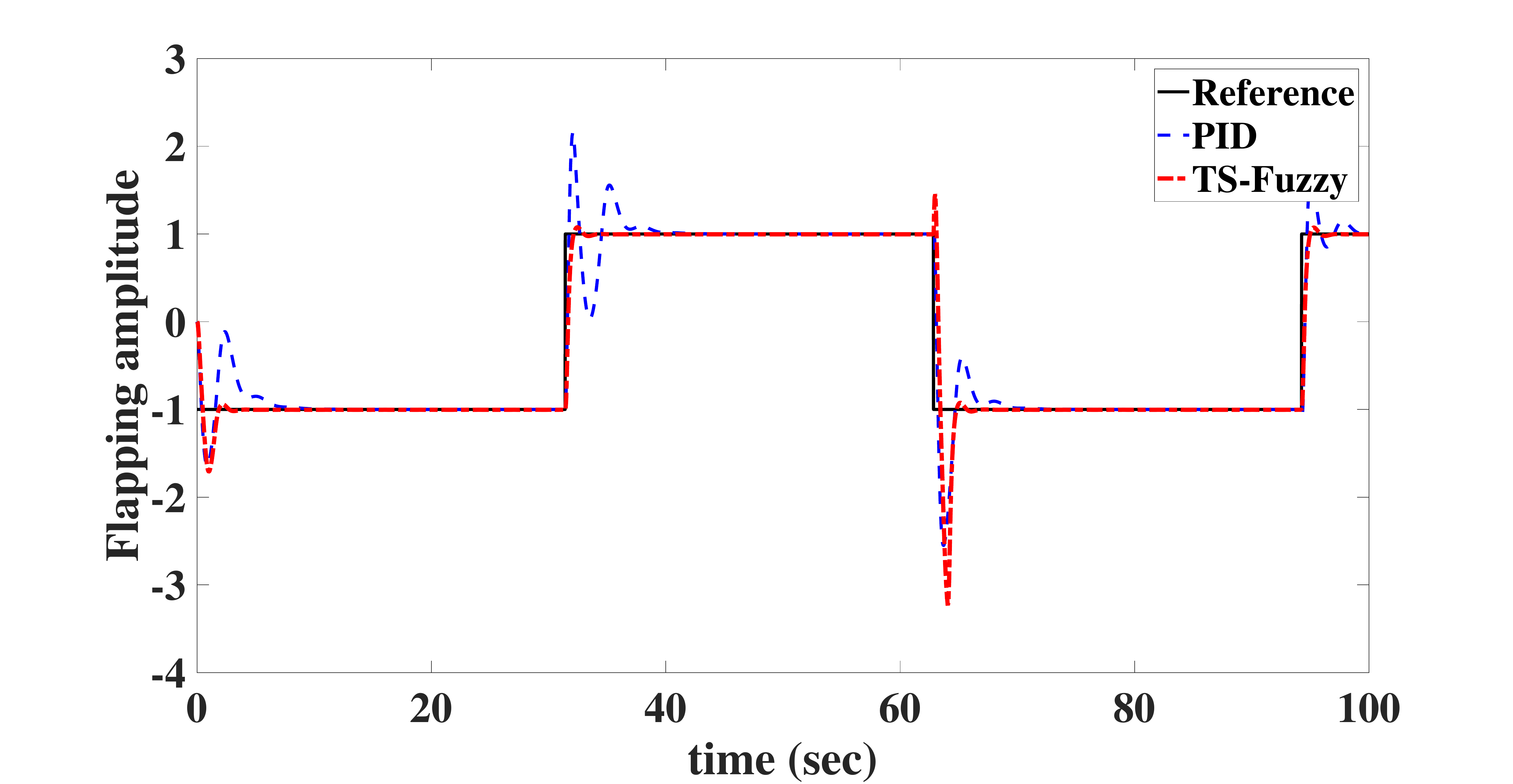}
	\caption{Altitude tracking performance of NIFW MAV controllers for a square wave trajectory}
	\label{fig:square}
\end{figure}

\begin{figure}[!t]
	\centering
	\includegraphics[width=3.5in]{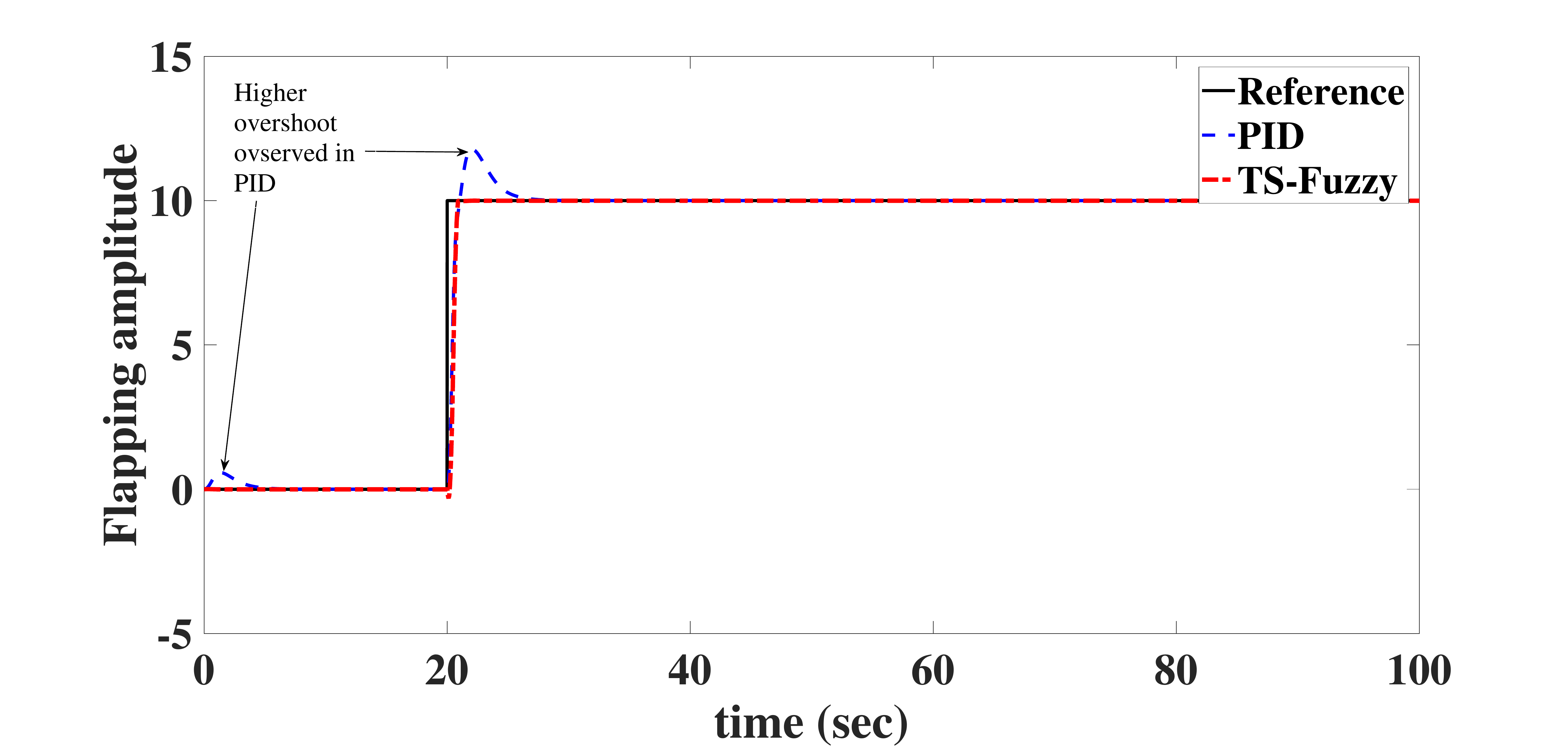}
	\caption{Altitude tracking performance of NIFW MAV controllers for a step wave trajectory changing amplitude from 0 m to 10 m after 20 sec.}
	\label{fig:step010}
\end{figure}

\begin{figure}[!t]
	\centering
	\includegraphics[width=3.5in]{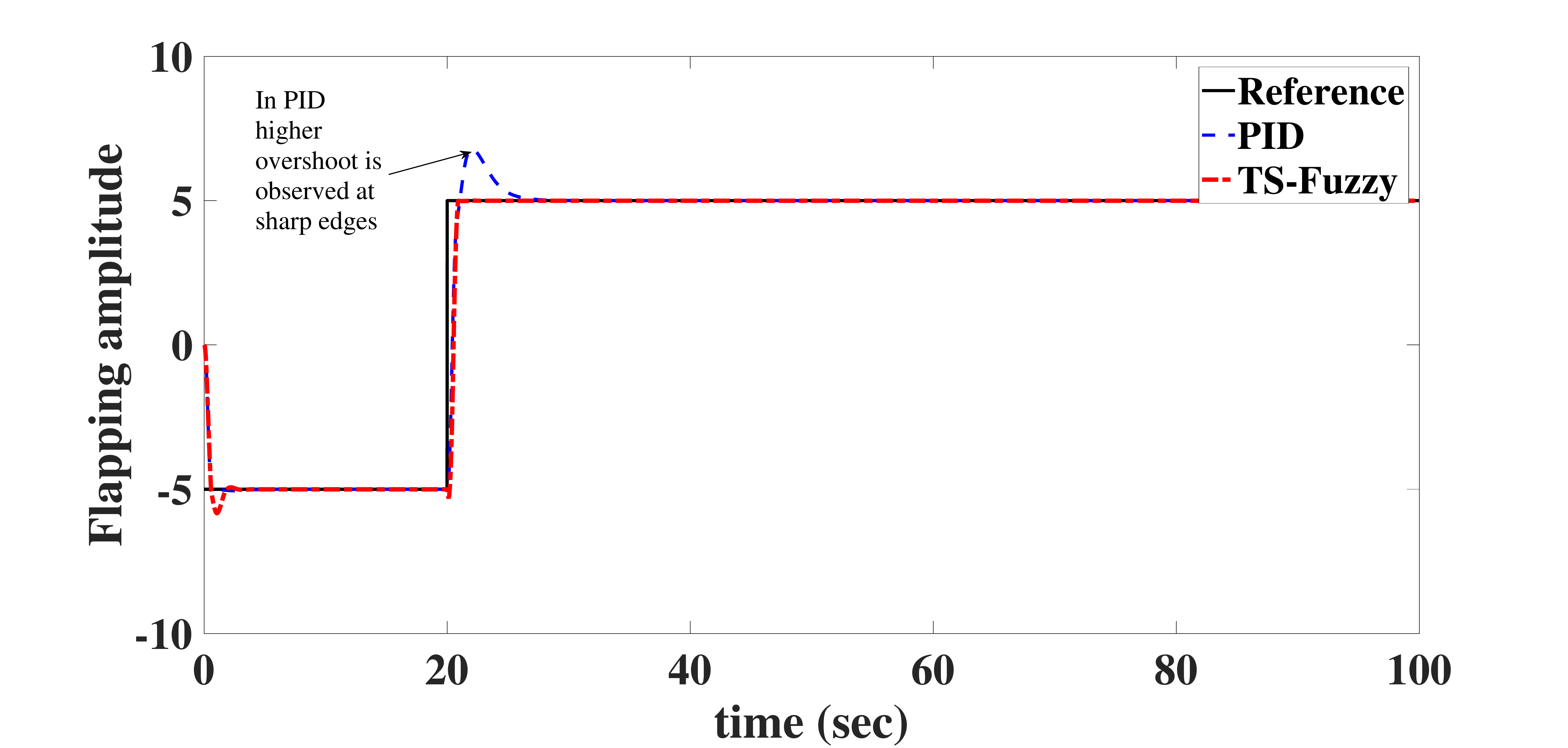}
	\caption{Altitude tracking performance of FW MAV controllers for a step wave trajectory changing amplitude from -5 m to 5 m after 20 sec.}
	\label{fig:step055}
\end{figure}

\begin{figure}[!t]
	\centering
	\includegraphics[width=3.5in]{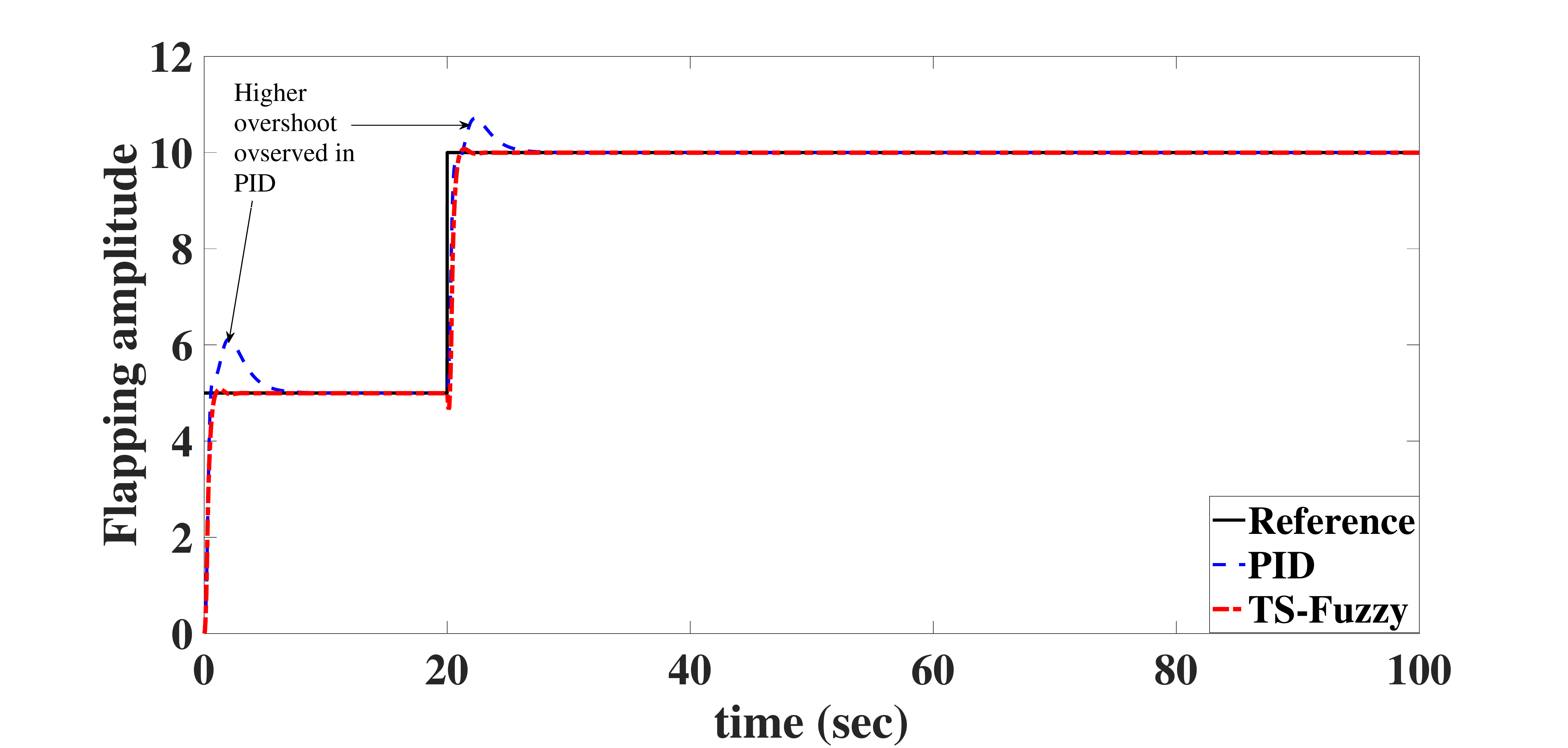}
	\caption{Altitude tracking performance of NIFW MAV controllers for a step wave trajectory changing amplitude from 5 m to 10 m after 20 sec.}
	\label{fig:step510}
\end{figure}

The adaptive TS fuzzy controller's performance is evaluated with respect to various reference signals and the results are compared with a  Proportional Integral Derivative (PID) controller. In this work, the considered desired trajectories for tracking altitude are constant height of 10 m, sinusoidal wave function with an amplitude of 1 m and frequency of 1 Hz, square wave function with an amplitude of 1 m and frequency of 0.1 Hz, and different step functions. At first, tracking performance of the controllers are observed for a trajectory of constant height. The results are observed in \figurename{~\ref{fig:cons_height}}, from where it is observed that the adaptive fuzzy controller performs better than the PID controller, besides a higher overshoot is observed from the PID controller at the beginning of tracking process. After that, performance is witnessed for a sine wave function reference and improved performance is observed from the fuzzy controller as exposed in  \figurename{~\ref{fig:sinusoidal}}. A square wave pulse with an amplitude of 1 m and frequency 0.1 Hz is also inserted into the closed loop system to observe the proposed adaptive fuzzy controller, and our proposed controller outperforms the PID controller as shown in \figurename{~\ref{fig:square}}. Finally, three different types of step functions presented as $Z_{b_d}(t)=10u(t-20)$, $Z_{b_d}(t)=5u(t)+5u(t-20)$,
$Z_{b_d}(t)=-5u(t)+10u(t-20)$ respectively are used as trajectories to check the proposed controller's performance. A satisfactory and better results are recorded in all cases as shown in \figurename{~\ref{fig:step010}}, \figurename{~\ref{fig:step055}}, and \figurename{~\ref{fig:step510}}. The root mean square error (RMSE) of both the PID and the proposed adaptive fuzzy controller for all the reference signals are tabulated in \tablename{~\ref{tab:Controllers}}, where lowest RMSE are obtained from the fuzzy controller in most of the cases.

\begin{table}[!t]
	\caption{Controllers performance (Measured RMSE)}\label{tab:Controllers}
	\centering{}%
	\begin{tabular}{|l|c|c|}
		\hline 
		\multirow{2}{*}{\textbf{Reference Signal}} & \multicolumn{2}{c|}{\textbf{RMSE}}\tabularnewline
		\cline{2-3} 
		& \textbf{PID} & \textbf{Adaptive Fuzzy}\tabularnewline
		\hline 
		Constant height & 0.6630 & 0.5693\tabularnewline
		\hline 
		Sinusoidal & 0.2096  & 0.0737\tabularnewline
		\hline 
		Square wave  & 0.2493 & 0.2039\tabularnewline
		\hline 
		Step 1  & 0.4000 & 0.4023\tabularnewline
		\hline 
		Step 2  & 0.6701 & 0.6266\tabularnewline
		\hline 
		Step 3  & 0.6695 & 0.7188\tabularnewline
		\hline 
	\end{tabular}
\end{table}

\section{Conclusion}
Acquiring exact numerical model of a highly nonlinear over-actuated complex system like NIFW MAV is difficult. Besides, the uncertainties are hard or sometimes impossible to incorporate. Recently researchers are trying to develop rule-based model free data-driven techniques using neural networks, fuzzy logic systems. Propelled by the various points of interest of those systems, in this work an FCM clustering based nonlinear fuzzy MIMO NIFW MAV model is developed, where the data has been gathered from a built-up NIFW MAV flight simulator. Moreover, the fuzzy clustering technique is utilized to develop an adaptive fuzzy controller and employed to control the MAVs altitude. To assess the controller's execution, the outcomes are contrasted with a PID controller with respect to constant height, sinusoidal wave, square wave, and three different step functions. In all cases, our developed adaptive fuzzy controller outperforms the PID controller, where the adaptive fuzzy is following the above-mentioned desired trajectories with an RMSE of only 0.5693, 0.0737, 0.2039, 0.4023, 0.6266 and 0.7188. In future, our adaptive fuzzy controller will be advanced into an evolving controller by utilizing an advanced learning machine algorithm and will be implemented in a NIFW MAV hardware.

\bibliographystyle{acm}
\bibliography{refAdfa}

\begin{thebibliography}{10}

\bibitem{Al-Mahasneh2017}
{\sc Al-Mahasneh, A.~A., Anavatti, S.~G., and Garratt, M.}
\newblock {Nonlinear Multi-Input Multi-Output System Identification using
  Neuro-Evolutionary Methods for a Quadcopter}, 2017.

\bibitem{almahasnehanavattigarratt2017}
{\sc Al-Mahasneh, A.~J., Anavatti, S., and Garratt, M.}
\newblock Altitude identification and intelligent control of a flapping wing
  micro aerial vehicle using modified generalized regression neural networks.
\newblock In {\em Computational Intelligence (IEEE SSCI), 2017 IEEE Symposium
  Series on\/} (2017), IEEE.

\bibitem{bezdek1980convergence}
{\sc Bezdek, J.~C.}
\newblock A convergence theorem for the fuzzy isodata clustering algorithms.
\newblock {\em IEEE transactions on pattern analysis and machine intelligence},
  1 (1980), 1--8.

\bibitem{clawson2016spiking}
{\sc Clawson, T.~S., Ferrari, S., Fuller, S.~B., and Wood, R.~J.}
\newblock Spiking neural network {(SNN)} control of a flapping insect-scale
  robot.
\newblock In {\em Decision and Control (CDC), 2016 IEEE 55th Conference on\/}
  (2016), IEEE, pp.~3381--3388.

\bibitem{couceiro2010modeling}
{\sc Couceiro, M.~S., Ferreira, N., and Machado, J.}
\newblock Modeling and control of a dragonfly-like robot.
\newblock {\em Journal of Control Science and Engineering 2010\/} (2010), 5.

\bibitem{couceiro2012hybrid}
{\sc Couceiro, M.~S., Ferreira, N.~M., and Machado, J.~T.}
\newblock Hybrid adaptive control of a dragonfly model.
\newblock {\em Communications in Nonlinear Science and Numerical Simulation
  17}, 2 (2012), 893--903.

\bibitem{du2016application}
{\sc Du, C.-p., Xu, J.-x., and Zheng, Y.}
\newblock Application of iterative learning tuning to a dragonfly-like flapping
  wing micro aerial vehicle.
\newblock In {\em Control and Decision Conference (CCDC), 2016 Chinese\/}
  (2016), IEEE, pp.~4215--4220.

\bibitem{ellington1999novel}
{\sc Ellington, C.~P.}
\newblock The novel aerodynamics of insect flight: applications to micro-air
  vehicles.
\newblock {\em Journal of Experimental Biology 202}, 23 (1999), 3439--3448.

\bibitem{ferdaus2017fuzzyclusteringFWMAV}
{\sc Ferdaus, M.~M., Anavatti, S.~G., Garratt, M.~A., and Pratama, M.}
\newblock {Fuzzy Clustering based Modelling and Adaptive Controlling of a
  Flapping Wing Micro Air Vehicle}.
\newblock In {\em Computational Intelligence (IEEE SSCI), 2017 IEEE Symposium
  Series on\/} (2017), IEEE, pp.~1914--1919.

\bibitem{ferdaus2017fuzzyclustering}
{\sc Ferdaus, M.~M., Anavatti, S.~G., Garratt, M.~A., and Pratama, M.}
\newblock {Fuzzy Clustering based Nonlinear System Identification and
  Controller Development of Pixhawk based Quadcopter}.
\newblock In {\em Advanced Computational Intelligence (ICACI), 2017 IEEE
  International Conference on\/} (2017), IEEE, pp.~223--230.

\bibitem{ferdaus2017McSIT2RFNN}
{\sc Ferdaus, M.~M., Anavatti, S.~G., Pratama, M., and Garratt, M.~A.}
\newblock {Online Identification of a Rotary Wing Unmanned Aerial Vehicle from
  Data Streams}.
\newblock 2017.

\bibitem{ferdaus2017GENEFISFWMAV}
{\sc Ferdaus, M.~M., Pratama, M., Anavatti, S.~G., and Garratt, M.~A.}
\newblock {Evolving Neuro-Fuzzy System based Online Identification of a
  Bio-inspired Flapping Wing Micro Aerial Vehicle}.
\newblock In {\em Computational Intelligence (IEEE SSCI), 2017 IEEE Symposium
  Series on\/} (2017), IEEE, pp.~2840--2847.

\bibitem{ferdaus2017Gcontroller}
{\sc Ferdaus, M.~M., Pratama, M., Anavatti, S.~G., and Garratt, M.~A.}
\newblock {Generic Evolving Self-Organizing Neuro-Fuzzy Control of Bio-inspired
  Unmanned Aerial Vehicles}.
\newblock 2018.

\bibitem{jang2003longitudinal}
{\sc Jang, J.~S., and Tomlin, C.}
\newblock Longitudinal stability augmentation system design for the
  {D}ragon{F}ly {UAV} using a single {GPS} receiver.
\newblock In {\em AIAA Guidance, Navigation, and Control Conference, AIAA\/}
  (2003), vol.~5592.

\bibitem{kok2015low}
{\sc Kok, J., and Chahl, J.}
\newblock A low-cost simulation platform for flapping wing {MAVs}.
\newblock In {\em SPIE Smart Structures and Materials+ Nondestructive
  Evaluation and Health Monitoring\/} (2015), International Society for Optics
  and Photonics, pp.~94290L--94290L.

\bibitem{kok2016design}
{\sc Kok, J.~M., and Chahl, J.}
\newblock Design and manufacture of a self-learning flapping wing-actuation
  system for a dragonfly-inspired {MAV}.
\newblock In {\em 54th AIAA Aerospace Sciences Meeting\/} (2016), p.~1744.

\bibitem{laurentevolution}
{\sc Laurent, J.-B. M. S.~D., and Druot, M.~T.}
\newblock Evolution of a neural network for the control of a flapping-wing
  animat.

\bibitem{lin2017optimal}
{\sc Lin, Y., Xu, Y., Chen, H., Bender, M.~J., Abbott, A.~L., and M{\"u}ller,
  R.}
\newblock {Optimal Threshold and LoG Based Feature Identification and Tracking
  of Bat Flapping Flight}.
\newblock In {\em Applications of Computer Vision (WACV), 2017 IEEE Winter
  Conference on\/} (2017), IEEE, pp.~418--426.

\bibitem{nguyen2014design}
{\sc Nguyen, Q.-V., Chan, W.~L., and Debiasi, M.}
\newblock Design, fabrication, and performance test of a hovering-based
  flapping-wing micro air vehicle capable of sustained and controlled flight.

\bibitem{nogar2017control}
{\sc Nogar, S.~M., Gogulapati, A., McNamara, J.~J., Serrani, A., Oppenheimer,
  M.~W., and Doman, D.~B.}
\newblock {Control-Oriented Modeling of Coupled Electromechanical-Aeroelastic
  Dynamics for Flapping-Wing Vehicles}.
\newblock {\em Journal of Guidance, Control, and Dynamics\/} (2017).

\bibitem{okamoto1996aerodynamic}
{\sc Okamoto, M., Yasuda, K., and Azuma, A.}
\newblock Aerodynamic characteristics of the wings and body of a dragonfly.
\newblock {\em Journal of Experimental Biology 199}, 2 (1996), 281--294.

\bibitem{oppenheimer2017wing}
{\sc Oppenheimer, M.~W., Sigthorsson, D.~O., Weintraub, I.~E., and Doman,
  D.~B.}
\newblock {Wing Design and Testing for a Tailless Flapping Wing Micro Air
  Vehicle}.
\newblock In {\em AIAA Guidance, Navigation, and Control Conference\/} (2017),
  p.~1271.

\bibitem{sane2003aerodynamics}
{\sc Sane, S.~P.}
\newblock The aerodynamics of insect flight.
\newblock {\em Journal of experimental biology 206}, 23 (2003), 4191--4208.

\bibitem{shyy2010recent}
{\sc Shyy, W., Aono, H., Chimakurthi, S.~K., Trizila, P., Kang, C.-K., Cesnik,
  C.~E., and Liu, H.}
\newblock Recent progress in flapping wing aerodynamics and aeroelasticity.
\newblock {\em Progress in Aerospace Sciences 46}, 7 (2010), 284--327.

\bibitem{shyy2007aerodynamics}
{\sc Shyy, W., Lian, Y., Tang, J., Viieru, D., and Liu, H.}
\newblock {\em Aerodynamics of low {R}eynolds number flyers}, vol.~22.
\newblock Cambridge University Press, 2007.

\bibitem{sudo1999study}
{\sc Sudo, S., Tsuyuki, K., Ikohagi, T., Ohta, F., Shida, S., and Tani, J.}
\newblock A study on the wing structure and flapping behavior of a dragonfly.
\newblock {\em JSME International Journal Series C Mechanical Systems, Machine
  Elements and Manufacturing 42}, 3 (1999), 721--729.

\bibitem{sun2010coupled}
{\sc Sun, J., Pan, C., Tong, J., and Zhang, J.}
\newblock Coupled model analysis of the structure and nano-mechanical
  properties of dragonfly wings.
\newblock {\em IET nanobiotechnology 4}, 1 (2010), 10--18.

\bibitem{tennekes2009simple}
{\sc Tennekes, H.}
\newblock {\em The simple science of flight: from insects to jumbo jets}.
\newblock MIT press, 2009.

\bibitem{wang2004role}
{\sc Wang, Z.~J.}
\newblock The role of drag in insect hovering.
\newblock {\em Journal of Experimental Biology 207}, 23 (2004), 4147--4155.

\bibitem{weng2013micro}
{\sc Weng, L., Xia, M., Hu, K., and Wang, A.}
\newblock {Micro Aerial Vehicle {(MAV)} Flapping Motion Control Using an Immune
  Network with Different Immune Factors}.
\newblock {\em International Journal of Advanced Robotic Systems 10}, 8 (2013),
  305.

\bibitem{willmott1997mechanics}
{\sc Willmott, A.~P., and Ellington, C.~P.}
\newblock The mechanics of flight in the hawkmoth manduca sexta. i. kinematics
  of hovering and forward flight.
\newblock {\em Journal of Experimental Biology 200}, 21 (1997), 2705--2722.

\bibitem{zhang2017review}
{\sc Zhang, C., and Rossi, C.}
\newblock A review of compliant transmission mechanisms for bio-inspired
  flapping-wing micro air vehicles.
\newblock {\em Bioinspiration \& biomimetics 12}, 2 (2017), 025005.

\bibitem{zhang2017resonance}
{\sc Zhang, J., and Deng, X.}
\newblock Resonance principle for the design of flapping wing micro air
  vehicles.
\newblock {\em IEEE Transactions on Robotics 33}, 1 (2017), 183--197.

\end{thebibliography}
\end{document}